\documentclass[journal]{IEEEtran}

\usepackage{cite}

\ifCLASSINFOpdf
   \usepackage[pdftex]{graphicx}

\else

\fi

\usepackage{amssymb}

\usepackage{amssymb, amsmath, amsfonts, amsthm}
\usepackage{fancyhdr}
\usepackage{color}
\usepackage{xcolor} 
\usepackage{graphicx}
\usepackage{epstopdf}
\usepackage{eurosym}
\usepackage{float}
\usepackage{verbatim} 
\usepackage{placeins}
\usepackage{caption}
\usepackage{subcaption} 
\usepackage{dblfloatfix}
\usepackage{epstopdf}
\usepackage[acronym,shortcuts]{glossaries}
\usepackage[margin=0.8in]{geometry}
\usepackage{multirow}
\usepackage{hyphenat}
\usepackage{cancel}

\makeglossaries
\newacronym{GFVSC}{\textcolor{black}{GFM-VSC}}{Grid-forming voltage source converter}

\begin{document}

\title{Revisiting angle stability in power systems with grid-forming power converters}
\author{R\'egulo E. \'Avila-Mart\'inez, Javier Renedo, Luis Rouco, \\ Aurelio Garcia-Cerrada, Lukas Sigrist
\thanks{This is an unabridged pre-print of the following paper (under review): \\ R. E. \'Avila-Mart\'inez, J. Renedo, L. Rouco, A. Garcia-Cerrada and Lukas Sigrist,  "Revisiting angle stability in power systems with grid-forming power converters," submitted to \emph{IEEE Power Engineering Letters}, Manuscript ID PESL-00471-2025.R1, pp. 1-4, 2025. \\ 
Internal ref: IIT-25-332WP. }
}
\maketitle
\begin{abstract}
This letter presents a comprehensive analysis of the stability phenomenon related to the ability of generators to remain in synchronism when subjected to small or large disturbances, in power systems with both synchronous machines and grid-forming voltage source converters (GFM-VSC). This phenomenon is associated with two stability classes in the IEEE/PES classification, namely, \emph{rotor-angle stability} (when involving synchronous machines) and \emph{slow-interaction converter-driven stability} (when involving power converters). However, this work shows that this phenomenon is fully characterised with the slow dynamics of the angle difference between the voltage sources connected to the power system, regardless of whether they are synchronous machines (with rotors) or GFM-VSCs. Therefore, we suggest using the term \emph{angle stability} to refer to this phenomenon, while \emph{slow-interaction converter-driven stability} should only include slow interactions of different nature involving power converters.
\end{abstract}
\begin{IEEEkeywords}
	Voltage source converter, VSC, grid forming, angle stability, transient stability, low-frequency oscillations
\end{IEEEkeywords}

\vspace{-0.4cm}
\section{Introduction}\label{sec.Intro}
\noindent Voltage source converters with grid-forming control (GFM-VSC) have been identified as a key technology to enable massive integration of renewable energy sources into power systems. One of the key challenges of power systems with large amounts of power converters is fully understanding their stability phenomena. Recently, IEEE/PES proposed an extended definition and classification of power system stability phenomena, taking into account key aspects of modern power systems with power converters~\cite{Nikos2021}. In the new classification, \emph{rotor-angle stability} remains unchanged with respect to the original definition~\cite{Kundur2004} and it is related to the ability of the synchronous machines to remain in synchronism when subject to small disturbances (electromechanical oscillations) or large disturbances (transient stability). However, during the last few years, several publications revealed that GFM-VSC power converters can also be subject to stability phenomenon related to their ability to remain in synchronism with the rest of the system when subject to small~\cite{Baruwa2021} or large disturbances~\cite{Pan2020}. The impact of key parameters is also analysed in~\cite{Pan2020}. Most of these publications use the term \emph{transient stability} to refer to the loss-of-synchronism phenomenon in GFM-VSC power converters, when subject to large disturbances~\cite{Pan2020}. Furthermore, previous work also showed that loss-of-synchronism phenomenon is also present in grid-following power converters (GFL-VSC) (see~\cite{Hu_GFL_transient_stab_2019}, for example). \color{black} However, the ability of GFM-VSCs to remain in synchronism when subject to small or large disturbances is already implicitly included by IEEE/PES~\cite{Nikos2021} into a class named \emph{converter-driven stability - slow interactions}. Therefore, the existing classification is not conducive to a
unified assessment of the power system’s overall synchronisation capability. 

It is well known that, both, synchronous machines and GFM-VSCs, can be represented as a voltage source (with a voltage magnitude and angle) behind a series impedance for power system stability studies. This letter illustrates that the ability of generators (synchronous machines or GFM-VSCs) to remain in synchronism can be described by analysing the slow dynamics of the angle difference between output voltages and, therefore, there is no need to address the stability issues of conventional synchronous machines and GFM-VSCs separately. Along this line, the letter proposes using the term \emph{angle stability} (instead of \emph{rotor-angle stability}), to refer to the stability phenomenon related to the capability of generators of the system to remain in synchronism under small or large disturbances, independently of the technology of these generators (synchronous generators or GFM-VSC power converters), as suggested in~\cite{Lindner2025}. The letter also proposes excluding \emph{angle stability} phenomenon from \emph{converter-driven stability - slow interactions}, in order to make the proposed definition consistent with IEEE/PES classification~\cite{Nikos2021}. 

It is important to highlight that the work in~\cite{Lindner2025} also proposed using the term \emph{angle stability} instead of \emph{rotor-angle stability} before. However, the work presented in this letter goes further and there are some differences in the approach. In~\cite{Lindner2025}, (a) a complete modification of IEEE/PES power-system-stability classification~\cite{Nikos2021} was proposed and (b) the use of the term \emph{angle stability} is explained, but detailed discussions, analyses and results supporting it were not presented in the paper. This letter, (a) just focuses on the definition of \emph{angle stability}, but it makes it consistent with the rest of stability classes of IEEE/PES~\cite{Nikos2021} and (b) it provides a comprehensive analysis to support the consistency of using the term \emph{angle stability} in power systems with grid-forming power converters. 

\vspace{-0.4cm}
\section{Results}\label{sec.results}
\begin{figure}[!htbp]
\begin{center}
\includegraphics[width=0.80\columnwidth]{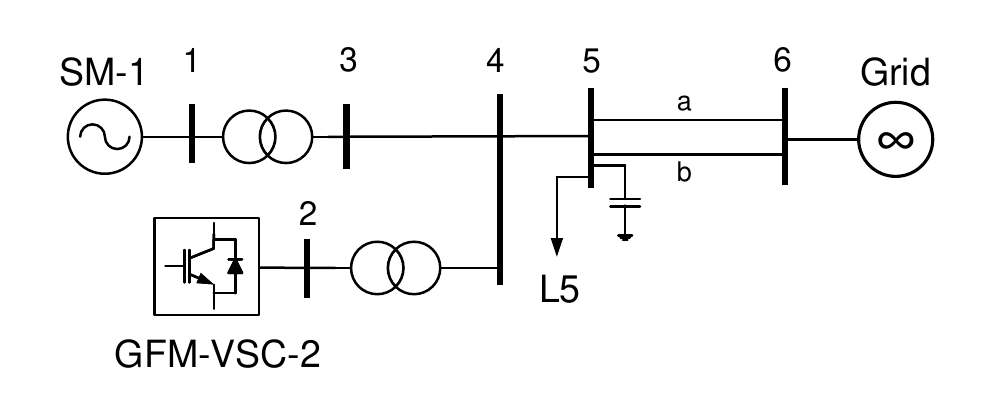}
\caption{Test system.}
\label{fig:Test_system_SM1_GFM2_IB_sld_v2}
\end{center}

\begin{center}
\includegraphics[width=0.45\textwidth]{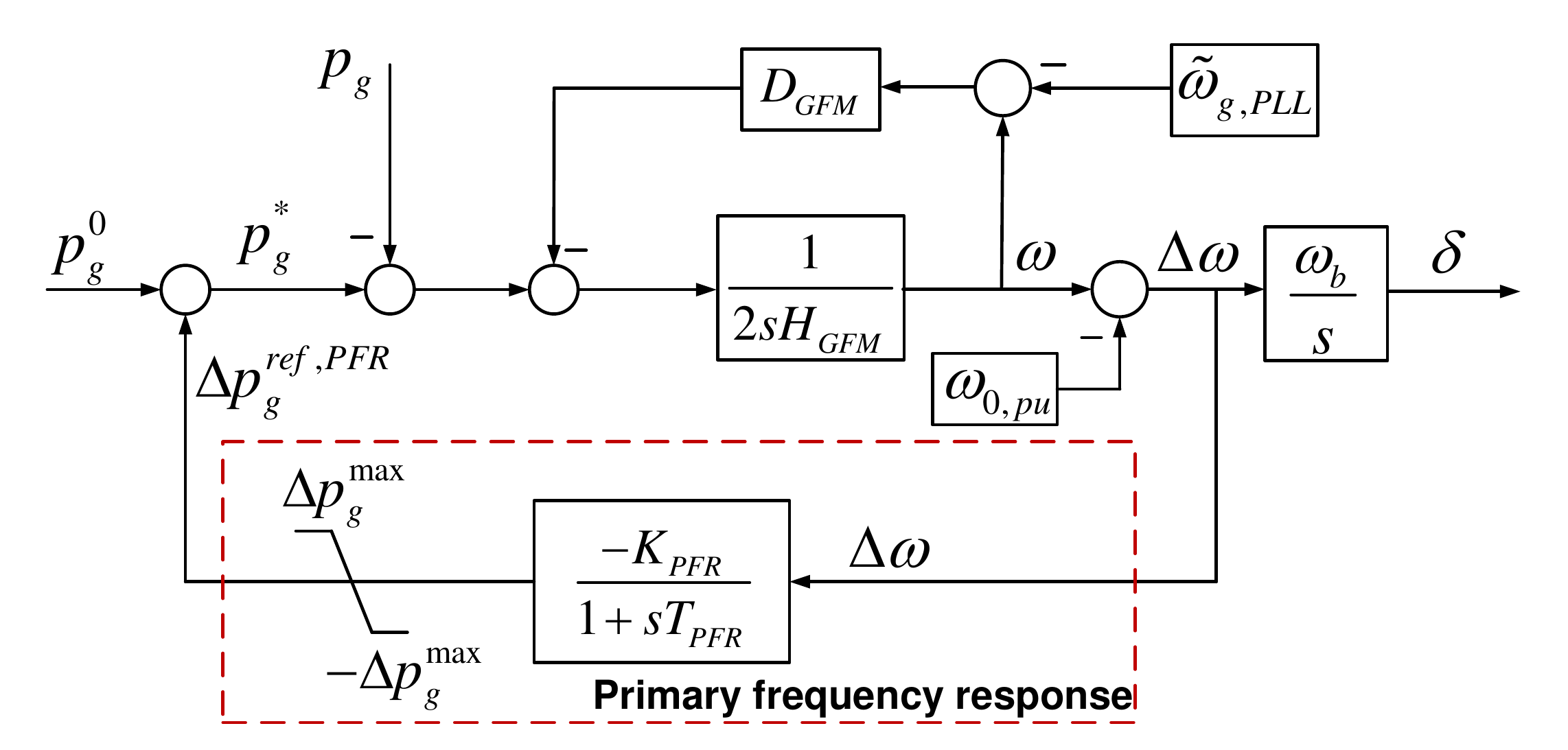}
\caption{GFM-VSC control scheme.}
\label{fig:GFM_VSC_VSM_PLL}
\end{center}
\end{figure}

\noindent Let us consider the power system in Fig.~\ref{fig:Test_system_SM1_GFM2_IB_sld_v2}, with a synchronous machine (SM-1) and a GFM-VSC-2 connected to an infinite grid. Each generator is rated to 900 MVA and is injecting $P_{g,1}^0=P_{g,2}^0=600$~MW at the initial operating point; and load $L_5$ \color{black} consumes 967~MW and 100~MVAr. GFM-VSC-2 has a current controller and a current limiter based on current-saturation algorithm (CSA) and it uses a grid-forming control scheme of the type Virtual Synchronous Machine (VSM) with frequency estimation at the connection point~\cite{DArco2015}, as shown in Fig.~\ref{fig:GFM_VSC_VSM_PLL}. All data of the test system used can be found in~\cite{ravila_test_system_2025}. Time-domain simulations were carried out using average electromagnetic-type (EMT) models in VSC\_Lib, which is an open-source toolbox based on Matlab + Simulink + Simscape~\cite{L2EP_VSC_GF_2020,MIGRATE_WP3_2018,Qoria2019a}.

\vspace{-0.3cm}
\subsection{Large disturbance}\label{sec.results_transient}
\begin{figure}[!htbp]
\begin{center}
\includegraphics[width=0.68\columnwidth]{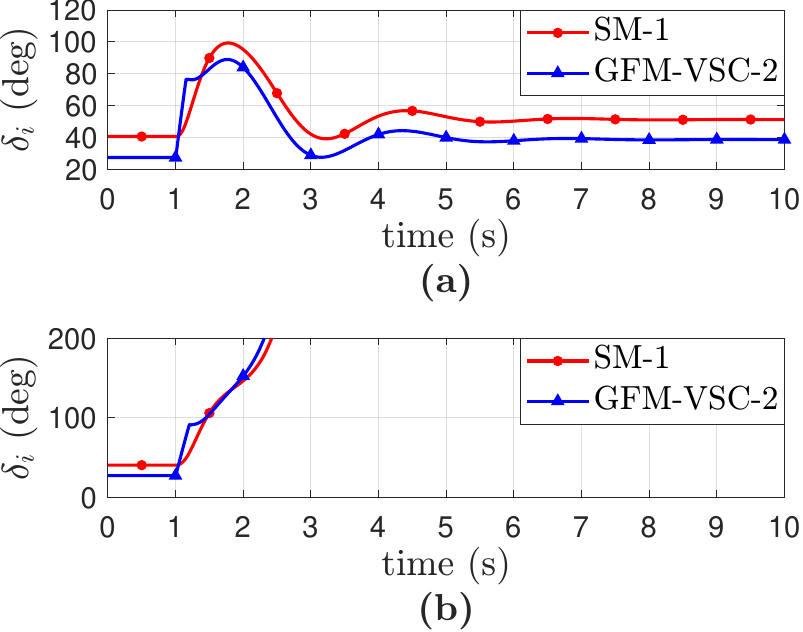}
\caption{Angles of SM-1 and GFM-VSC-2. (a) Fault 1 cleared after 150~ms and (b) Fault 1 cleared after 200~ms.}
\label{fig:Study_1_TS_CaseA_Fault1_Angles}
\end{center}
\end{figure}

\noindent A three-phase-to-ground short circuit was applied to line 5-6a (close to bus 5) at $t=1$~s, and it was cleared by disconnecting the line after 150~ms (Fault 1, for short). Fig.~\ref{fig:Study_1_TS_CaseA_Fault1_Angles}-a shows the  rotor angle of SM-1 and the angle of the modulated voltage of GFM-VSC-2, both with respect to the infinite grid. The system is stable and both generators remain in synchronism.

For a clearing time of Fault 1 of 200~ms, the system is unstable and both generators SM-1 and GFM-VSC-2 lose synchronism (Fig.~\ref{fig:Study_1_TS_CaseA_Fault1_Angles}-b). In fact, the Critical Clearing Time~(CCT) of Fault 1 is 180~ms. According to~\cite{Nikos2021}, the loss-of-synchronism of SM-1 belongs to \emph{large-disturbance rotor-angle stability} (transient stability), while the loss-of-synchronism of GFM-VSC-2 belongs to \emph{slow-interaction converter-driven stability}.

As a conclusion, the overall instability observed in Fig.~\ref{fig:Study_1_TS_CaseA_Fault1_Angles}-b is related to a single phenomenon: the ability of synchronous machine SM-1 and grid-forming power converter GFM-VSC-2 to remain in synchronism after large disturbances.

\vspace{-0.3cm}
\subsection{Small disturbance}\label{sec.results_osc}
\begin{figure}[!htbp]
\begin{center}
\includegraphics[width=0.68\columnwidth]{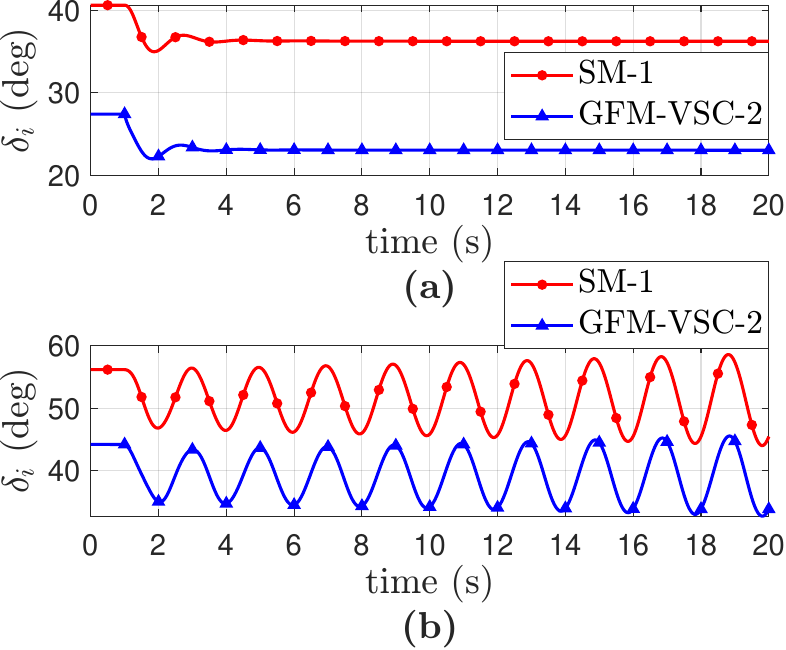}
\caption{Angles of SM-1 and GFM-VSC-2. Load reduction of 10~\% in load L5. (a) Case A and (b) Case B.}
\label{fig:Study_2_Angle_SSA_CaseAB_LoadChng_10p_Angles}
\end{center}
\end{figure}

\noindent A load reduction of 10~\% was applied to load $L_5$ at $t=1$~s (a small disturbance). Fig.~\ref{fig:Study_2_Angle_SSA_CaseAB_LoadChng_10p_Angles}-a shows the  rotor angle of SM-1 and the angle of GFM-VSC-2. The system is stable and the oscillations observed in the angles of both generators are well damped (case A, for short).

In a second case study (Case B), to obtain a critical case in terms of oscillatory behaviour, the active-power injections of the generators were increased to $P_{g,1}^0=P_{g,2}^0=700$~MW, the Power System Stabilizer (PSS) of SM-1 was deactivated and the damping coefficient of GFM-VSC-2 was reduced from $D_{GFM}=193$~pu to $D_{GFM}=20$~pu, artificially. The same small disturbance as before was simulated now (load reduction of 10~\% in load $L_5$ at $t=1$~s). Fig.~\ref{fig:Study_2_Angle_SSA_CaseAB_LoadChng_10p_Angles}-b shows that the system is unstable now and undamped 0.51~Hz low-frequency oscillations can be observed in, both, the rotor angle of SM-1 and the angle of GFM-VSC-2, with respect to the infinite grid. Both, SM-1 and GFM-VSC-2 oscillate against the infinite grid. According to~\cite{Nikos2021}, the instability of SM-1 belongs to \emph{small-disturbance rotor-angle stability}, while the instability in GFM-VSC-2 belongs to \emph{slow-interaction converter-driven stability}.

As a conclusion, the overall instability observed in Fig.~\ref{fig:Study_2_Angle_SSA_CaseAB_LoadChng_10p_Angles}-b is related to a single phenomenon: the ability of generators SM-1 and GFM-VSC-2 to remain in synchronism after small disturbances. 

\vspace{-0.3cm}
\section{Why not just angle stability?}\label{sec.angle_stab_def}
\noindent According to the results of previous section, the following limitations in current stability classification of IEEE/PES~\cite{Nikos2021} have been identified: 
\begin{itemize}
\item The large-disturbance instability phenomenon of Fig.~\ref{fig:Study_1_TS_CaseA_Fault1_Angles}-(b) involves two stability phenomena (\emph{large-disturbance rotor-angle stability} and \emph{slow-interaction converter-driven stability}), not one.
\item The small-disturbance instability phenomenon of Fig.~\ref{fig:Study_2_Angle_SSA_CaseAB_LoadChng_10p_Angles}-(b) also involves two stability phenomena (\emph{small-disturbance rotor-angle stability} and \emph{slow-interaction converter-driven stability}), not one.
\item \emph{Slow-interaction converter-driven stability}, which includes stability phenomenon related to the ability of GFM-VSCs to remain in synchronism, does not distinguish between small- or large-disturbance stability, which is relevant in this context.
\item \emph{Slow-interaction converter-driven stability} includes stability phenomenon related to the ability of GFM-VSCs to remain in synchronism, but this class also includes converter-driven interactions of very different nature.
\end{itemize}
The results of previous section also illustrate that stability phenomena related to synchronism of generators in power systems with synchronous machines and GFM-VSC power converters could be understood as a single phenomenon, related to the stability of the angle of the voltage source representing each device, independently of its technology. 

Along this line, this letter proposes the following modifications to the IEEE/PES classification~\cite{Nikos2021} (in Table~\ref{tab:angle_stab_proposed_definition}). 

\begin{table}[h]
\centering
\caption{Proposed modifications to the IEEE/PES power system stability classification~\cite{Nikos2021}. Proposed modifications in bold. }
\begin{tabular}{|p{2.50cm}|p{5.25cm}|}
\hline
 \color{blue} \textbf{Stability class} \color{black} & \color{blue} \textbf{Definition} \color{black} \\ \hline
\color{blue} \textbf{Angle stability}  $ $ $ $ $ $ $ $ (\emph{angle stability} is used instead of \emph{rotor-angle stability}) \color{black} & \color{blue} \emph{The ability of the interconnected synchronous machines} \textbf{and power-electronics-based} devices \emph{in a power system to remain in synchronism under normal operating conditions and to regain synchronism after being subjected to a small or large disturbance}. \color{black} \\ \hline
\color{blue} Converter-driven stability - slow interactions \color{black} & \color{blue} \emph{Instabilities that involve system-wide instabilities driven by slow dynamic interactions of the control systems of power-electronic-based devices with slow response components of the power system}, \textbf{but excluding angle-stability phenomenon.} \color{black} \\ \hline
\end{tabular}
\label{tab:angle_stab_proposed_definition}
\end{table}
\FloatBarrier
This work does not propose any other modification related to the rest of stability phenomena of~\cite{Nikos2021}. 

In the proposed definition, \emph{angle stability} is used instead of \emph{rotor-angle stability}, as suggested in~\cite{Lindner2025}. \emph{Large-disturbance angle stability} (transient stability) is a non-linear phenomenon and it should be assessed by non-linear time-domain simulation or analysis based on direct methods. Since the dynamics of interest are slow, electromechanical-type models could be used (also known as Root-Mean-Square (RMS) models). \emph{Small-disturbance angle stability} does not include just electromechanical oscillations any more: low-frequency oscillations related to this phenomenon could be called \emph{Angle Electromechanical/Power-Electronics (AEPE) oscillations}. AEPE oscillations can be classified into local or inter-area oscillations, analogously to traditional electromechanical oscillations. \emph{Small-disturbance angle stability} can be assessed by small-signal stability analysis (eigenvalue and frequency-domain methods), using a RMS-type linearised model of the power system around an operating point.

With the modifications proposed in this letter, \emph{slow-interaction converter-driven stability}~\cite{Nikos2021} would include slow interactions involving converters' controllers, but excluding the ability of the power converters to remain in synchronism when subject disturbances. This category is wide and it involves interactions of different nature. An example of \emph{slow-interaction converter-driven stability} is the 0.1~Hz oscillation reported in~\cite{Fan2023}, which is related to control interactions of power converters when connected to the grid, but it is not related to the ability of synchronous machines or power converters to remain in synchronism when subject to disturbances.

\vspace{-0.4cm}
\section{Conclusion}\label{sec.conclusion}
\noindent This letter illustrates the paradigm that, currently, in a power system with both,  synchronous generators and GFM-VSC power converters, the stability phenomenon related to the ability of those devices to remain in synchronism when subject to small or large disturbances would be classified into two different categories, namely, \emph{rotor-angle stability} and \emph{slow-interaction converter-driven stability}, although, in fact, both phenomena involve the same type of system
variables (angles of the voltage sources representing those generators). Therefore, we propose the use of the name of \emph{angle stability} (instead of \emph{rotor-angle stability}) to refer to this phenomenon, regardless the type of  devices involved. This change would leave the current IEEE/PES \emph{slow-interaction converter-driven stability} category for any other slow converter-related phenomenon, not related to the angle of its output voltage and its ability to maintain synchronism. An interesting research line would be to analyse in detail angle stability of different applications of GFM/GFL power converters and taking into account the primary energy source, identifying the key aspects related to each application. 

The modifications proposed in this letter are relevant to add consistency to power system stability classification. However, the proposed framework is also relevant from a problem-solving perspective. For example, \emph{Slow-interaction converter-driven stability} includes interactions of very different kind, and mitigation methods would depend on the particular phenomenon observed in a certain event. Nevertheless, by including the ability of GFM-VSCs to maintain synchronism into a general \emph{angle stability} class, it allows to use well-established mitigation methods for events related to this phenomenon, such as (a) supplementary controllers in power converters to improve large-disturbance angle stability (transient stability), or (b) power oscillation damping (POD) controllers in power converters and power system stabilisers (PSS) in synchronous machines to improve small-disturbance angle stability (AEPE oscillations); among others. 

Finally, although this paper suggests using the generic term \emph{angle stability}, it is important to understand the physics behind the instability mechanisms involved in each device. Angle stability phenomenon in synchronous machines is related to their electromechanical behaviour, while angle stability phenomenon in GFM-VSCs is related to a control algorithm (control structure and parameters).  

\section{Acknowledgements}\label{sec.acknowledgements}
\noindent Work supported by the Spanish Government under a research project ref. PRE2019-088084, RETOS Project  Ref. RTI2018-098865-B-C31 (MCI/AEI/FEDER, UE) and Project PID2021-125628OB-C21 (MICIU/AEI /10.13039/501100011033 and FEDER, EU);  and by Madrid Regional Government under PROMINT-CM Project  Ref. S2018/EMT-4366.  

\section{Authors' information}\label{sec.authors_infromation}
\noindent Régulo E. Ávila-Martínez, Luis Rouco, Aurelio Garcia-Cerrada, and Lukas Sigrist are with the Instituto de Investigación Tecnológica~(IIT), ETSI
ICAI, Universidad Pontificia Comillas, Madrid, Spain (e-mail: regulo.avila@iit.comillas.edu; luis.rouco@iit.comillas.edu; aurelio@comillas.edu;
lukas.sigrist@iit.comillas.edu). Javier Renedo is with the ETSI ICAI, Universidad Pontificia Comillas, Madrid, Spain (e-mail: javier.renedo@ieee.org). 



\end{document}